\documentclass[12pt]{article}

\usepackage{geometry}
\usepackage{amssymb}
\usepackage{epsf}

\newcommand{\bra}[1]{\langle {#1} |}
\newcommand{\ket}[1]{| {#1} \rangle}
\newcommand{\vecr}{{\bf r}}
\newcommand{\Journal}[4]{{#1} {\bf #2}, #3 (#4)}
\def\NPA{{\em Nucl. Phys.} A}
\def\PRC{{\em Phys. Rev.} C}
\def\JCP{{\em J. Chem. Phys.}}
\def\MP{{\em Mol. Phys.}}
\def\JPSJS{{\em J. Phys. Soc. Japan Suppl.}}
\def\JComP{{\em J. Comput. Phys.}}

\begin{document}

\title{Resonance and Continuum States in Weakly Bound Systems}
\author{Takashi~Nakatsukasa\footnote{
Present address: Physics Department, Tohoku University,
Sendai 980-8578, Japan}
\ and Kazuhiro~Yabana$^\dagger$\\
\itshape RI Beam Science Laboratory, RIKEN, Wako 351-0198, Japan\\
\itshape $^\dagger$Institute of Physics, University of Tsukuba,
Tsukuba 305-8571, Japan}
\maketitle

\abstract{
Linear response theories in the continuum capable of
describing continuum spectra and dynamical correlations
are presented.
Our formulation is essentially the same as the continuum random-phase
approximation (RPA) but
suitable for uniform grid representation in the three-dimensional (3D)
Cartesian coordinate assuming no spatial symmetry.
Effects of the continuum are taken into account either by solving equations
iteratively with a retarded Green's function or by an absorbing boundary
condition.
The methods are applied to giant resonances in a
deformed nucleus $^{12}$C.
}

\section{Introduction}
\label{sec: intro}
Recent developments in radioactive beam techniques enable us to access
nuclei near and beyond drip-lines.
The drip-line nuclei are weakly bound finite fermion systems.
One would naturally expect that the continuum should be taken into
account explicitly in description of their structure and reaction.

Bound solutions of the Schr\"odinger equation for the Hamiltonian
$H=T+V$
\begin{equation}
\left( E-T \right) \Psi = V\Psi ,
\end{equation}
can be formally written as
\begin{equation}
\label{LS_bound}
\Psi = \frac{1}{E-T} V \Psi.
\end{equation}
Here, a Green's function $(E-T)^{-1}$
can be uniquely defined
since the operator $E-T$ is a negative-definite operator.
Bound states are characterized by discrete spectra,
$E_n$ $(n=0,1,2,\cdots)$.

On the other hand, for continuum states, since the $E-T$ has zero
eigenvalues, we need to modify Eq. (\ref{LS_bound}) into
the Lippmann-Schwinger equation.
First, we need to add a zero-eigenvalue solution of the $E-T$, $\Phi_0$.
Furthermore, in order to uniquely define the Green's function
$(E-T)^{-1}$,
it is necessary to specify a boundary condition.
For usual physical situations,
it is natural to adopt an outgoing boundary condition (OBC)
for scattering waves,
\begin{equation}
\label{LS_scattering}
\Psi = \Phi_0+\frac{1}{E-T+i\eta} V \Psi.
\end{equation}
For two-body systems, one may easily separate relative coordinates
from those of the center of mass.
Then, the OBC becomes rather trivial
because this is essentially a problem of a single degree of freedom.
For many-body systems, however, it can be very complicated to
settle the OBC for many degrees of freedom.
We focus our discussion upon treatment of this problem.

The Skyrme Hartree-Fock (HF) theory in the three-dimensional (3D)
Cartesian coordinate representation\cite{Flo78}
has been extensively applied to
study of ground-state properties of relatively heavy nuclei.
To investigate their excited states,
a straightforward extension is either the time-dependent Hartree-Fock (TDHF) 
or the linear response calculation based on the HF ground state.
For spherical nuclei, since the HF problem has a single degree of freedom
(the radial coordinate from the center of mass),
one can easily treat the boundary condition.
Therefore,
the inclusion of the single-particle continuum for particle-hole (p-h) 
excitations has been achieved by using a method, so called, continuum
random-phase approximation (RPA)\cite{SB75}.
The continuum RPA combined with the Skyrme Hartree-Fock (HF) theory
has been extensively utilized to study giant resonances (GR)
in spherical nuclei\cite{GS81,HSZ96}.
However, it is not trivial to extend the method to deformed nuclei.

Since the Hamiltonian in the Skyrme HF theory is almost diagonal in
coordinate representation, a grid representation in the coordinate space
provides an economical description.
The main issue is then how to incorporate the boundary condition
in the 3D uniform grid representation.
We shall present prescriptions to treat the full
three dimensional (3D) continuum in the RPA level.

Recently, we have investigated photoabsorption cross sections of
small molecules with essentially the same techniques\cite{NY01}.
In this paper, we report the applications to nuclear giant resonances
with a simplified Skyrme Hamiltonian.

\section{Linear response in the continuum}
\label{sec:boundary_cond}
\subsection{Outgoing boundary condition (OBC) in the 3D space}
\label{subsec: OBC}

In the linear response theory based on the Skyrme HF,
exact treatment of the single-particle continuum is possible
utilizing a single-particle Green's function.
For spherical systems, the Green's function can easily
be constructed by making a multipole expansion and discretizing the
radial coordinate.
This is an essential part of the continuum RPA method.
In this section we present a method to construct
a single-particle Green's function in the 3D grid representation
for a system without any spatial symmetry.

The linear response theory is formulated most conveniently using a
retarded density-density correlation function\cite{FW71}.
\begin{eqnarray}
\Pi(\vecr,\vecr';\omega)&=& \int dt e^{i\omega t-\eta t}
                            \Pi(\vecr,t;\vecr',0) ,\\
\label{Pi_def}
i \Pi(\vecr,t;\vecr',t') &=& \theta(t-t')
  \bra{0} [ \hat{\rho}(\vecr,t),\hat{\rho}(\vecr',t') ] \ket{0} .
\end{eqnarray}
In the RPA of retaining only ring diagrams,
$\Pi(\vecr,\vecr';\omega)$ is constructed from 
independent-particle density-density correlation function,
$\Pi_0$, which is defined by
identifying the state $\ket{0}$ in Eq. (\ref{Pi_def})
with the unperturbed HF ground state and assuming that
the density operator is evolved in time
with the static HF Hamiltonian.
Im$\Pi(\omega)$ is proportional to the excitation probability from
the ground state to a state of excitation energy $\omega$.

The $\Pi_0$ is written in a form\cite{SB75}
\begin{equation}
\label{Pi_0}
\Pi_0(\vecr,\vecr';\omega)=
 \sum_{i=1}^A
 \phi_i(\vecr)\left\{
     \left(G^{(+)}(\vecr,\vecr';(\epsilon_i-\omega)^*)\right)^*
   + G^{(+)}(\vecr,\vecr';\epsilon_i+\omega) \right\}
   \phi_i(\vecr') ,
\end{equation}
assuming that the occupied states have real wave functions $\phi_i$
and eigenenergies $\epsilon_i$.
The $G^{(+)}$ is a single-particle Green's function with OBC,
defined by
\begin{equation}
\label{G_sp}
G^{(+)}(\vecr,\vecr';E)
= \bra{\vecr}
\left( E + \frac{1}{2m}\nabla^2 - V_{\rm HF}(\vecr) + i\eta \right)^{-1}
\ket{\vecr'} ,
\end{equation}
where $V_{\rm HF}$ is the HF potential which is not necessarily spherical.
It is difficult to explicitly construct the $G^{(+)}(\vecr,\vecr';E)$
in the 3D space.
For bound states ($E<0$), instead,
we can calculate actions of $G^{(+)}(E)$ on a state $\Phi$ by
solving a differential equation
\begin{equation}
\label{Sternheimer}
\left(E + \frac{1}{2m}\nabla^2 - V_{\rm HF}(\vecr)\right) \Psi(\vecr)
    = \Phi(\vecr) ,
\end{equation}
where $\Psi=G^{(+)}(E)\Phi$.
However, as we have mentioned in Sec. \ref{sec: intro},
we have to specify the OBC for the continuum states with $E>0$.

For this task,
we decompose the HF potential
into
a spherical part $V_0$ and
 a short-range deformed part $V_{\rm D}$ ($V_{\rm HF}=V_0+V_{\rm D}$).
We first construct a Green's function,
$G_0^{(+)}(E)\equiv (E-H_0+i\eta)^{-1}$,
for the spherical Hamiltonian, $H_0=-1/2m\nabla^2 + V_0$,
by using the usual multipole expansion technique\cite{SB75}.
Next, we solve an equation
\begin{equation}
\label{Psi_eq}
(1 -G_0^{(+)}(E) V_{\rm D}) \Psi = G_0^{(+)}(E) \Phi ,
\end{equation}
to calculate
$\Psi=G^{(+)}(E)\Phi=(E-H_0-V_{\rm D}+i\eta)^{-1}\Phi$.
Eq. (\ref{Psi_eq}) can be derived from 
an identity for the Green's function
\begin{equation}
G^{(+)}(E)=G_0^{(+)}(E) + G_0^{(+)}(E) V_{\rm D} G^{(+)}(E) .
\end{equation}
In this way, we fix an outgoing asymptotic behavior of the
Green's function $G^{(+)}(E)$.
The details of numerical procedure is described in our recent
paper\cite{NY01}.

\subsection{Absorbing boundary condition (ABC) equivalent to OBC}

In this section, we present another method to simulate the OBC,
which we call ``absorbing boundary condition'' (ABC)\cite{HM78,KK86}.
The Green's function with OBC is written as Eq. (\ref{G_sp})
where $+i\eta$ is an infinitesimal imaginary quantity ($\eta>0$).
Now, we allow this imaginary part to depend on coordinate
and to be finite, $+iW(\vecr)$.
$W(\vecr)$ is taken to be positive far outside the system (at large $r$)
and zero elsewhere.
This means that, for $E>0$,  the wave number $k$ has a positive
imaginary part $+i\gamma(\vecr)$ at large $r$.
The outgoing wave defines its asymptotic behavior as
\begin{equation}
\psi^{(+)} \sim f(\Omega) \frac{e^{ikr}}{r},
 \quad\quad\mbox{at } r\rightarrow \infty.
\end{equation}
Thus, with the complex potential $-iW(\vecr)$,
the outgoing wave is going to damp as
\begin{equation}
\psi^{(+)} \sim f(\Omega) \frac{e^{ikr-\gamma r}}{r},
 \quad\quad\mbox{at } r\rightarrow \infty,
\end{equation}
while the incoming wave is going to diverge.
Therefore, if we impose the vanishing boundary condition at large $r$, 
only outgoing waves are allowed.
Since the complex potential takes care of the boundary condition,
we can directly solve Eq. (\ref{Sternheimer}) with
$\Psi|_{\rm boundary}=0$, instead of constructing the OBC by solving
Eq. (\ref{Psi_eq}).
In this sense, the ABC is easier to handle than the OBC in
Sec \ref{subsec: OBC}.

The remaining task is how to construct
a good absorbing boundary potential.
When an outgoing wave hits the complex potential $-iW(\vecr)$,
a part of the wave is absorbed and another part is reflected back.
Since the reflected part causes significant inaccuracies,
the complex boundary potential must be strong enough
to absorb the whole wave and simultaneously gentle enough
to minimize the reflection.
We adopt an absorbing potential of linear dependence on
the radial coordinate,
\begin{equation}
W(r) = 
\cases{
0 ,                         &  \mbox{for $r < R$,} \cr
W_0 \frac{r-R}{\Delta r},   &  \mbox{for $R < r < R+\Delta r$.}\cr
}
\end{equation}
In order to minimize the spurious reflection,
the height $W_0$ and width $\Delta r$ should satisfy
a condition\cite{Chi91,NY01}
\begin{equation}
\label{good_absorber}
20\frac{E^{1/2}}{\Delta r \sqrt{8m}} < \left| W_0 \right|
< \frac{1}{10} \Delta r \sqrt{8m} E^{3/2} .
\end{equation}

An advantage of the ABC is simplicity of its numerical calculation.
We may solve scattering problems in the same way as we do for bound states.
Another advantage is that
the ABC easily implements a real-time calculation\cite{HM78,NY01}.
Instead of taking an energy representation, we may calculate time evolution
of the TDHF states directly.
We have demonstrated that the real-time method with the ABC can
properly take account of the continuum in the linear response
calculations\cite{NY01}.
In real-time calculations, it is very difficult to handle the OBC explicitly.
The disadvantage is that the ABC cannot handle low-energy escaping particles
and a long-range Coulomb potential properly.
Therefore, the outgoing protons feel a finite-range Coulomb potential,
$Ze/r$ only at $r<R$.

\section{Giant resonances in the continuum}
\label{sec: GR}
\subsection{Giant monopole resonances}

We apply the methods to isoscalar giant resonances (GMR) in $^{12}$C.
We use an energy functional given by the ``BKN Hamiltonian''\cite{BKN76}:
\begin{eqnarray}
E[\rho] &=& \int d\vecr \left[
  \frac{1}{2m} \sum_{i=1}^A\left|\nabla\phi_i(\vecr)\right|^2
 + \frac{3}{8}t_0\rho(\vecr)^2
  +\frac{1}{16}t_3\rho(\vecr)^3 \right]\nonumber\\
 &&+ \frac{1}{2} V_0 a \int d\vecr \int d\vecr'
   \rho(\vecr) \frac{\exp(-|\vecr-\vecr'|/a)}{|\vecr-\vecr'|}\rho(\vecr')
   \nonumber\\
 &&+ \frac{e^2}{8} \int d\vecr\int d\vecr'
   \rho(\vecr) \frac{1}{|\vecr-\vecr'|}\rho(\vecr') ,
\end{eqnarray}
where spin-isospin degeneracy (each nucleon with a charge $e/2$) is
assumed, $\rho(\vecr)=4\sum_i |\phi_i(\vecr) |^2$.

\begin{figure}[t]
\epsfxsize=0.8\textwidth
\centerline{\epsfbox{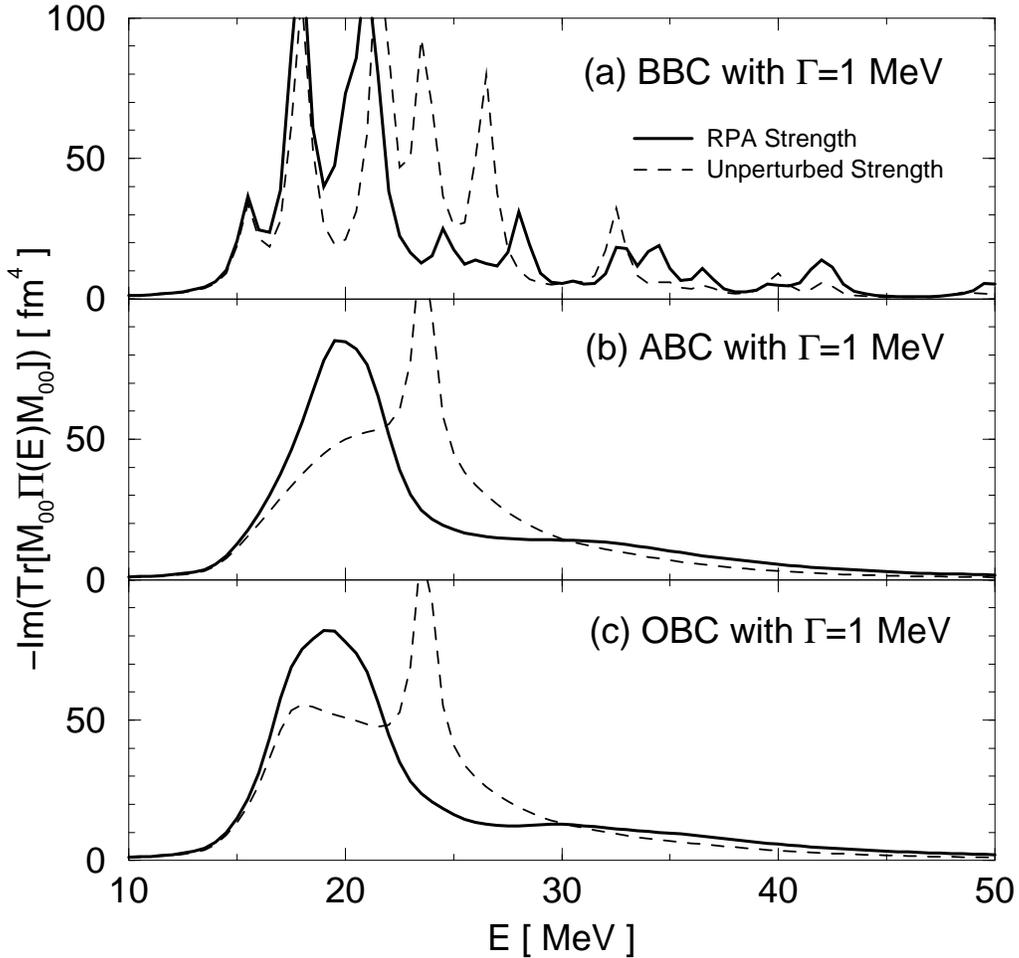}}
\vspace*{-10pt}
\centerline{
\begin{minipage}{0.8\textwidth}
\caption{
Monopole strength distribution,
$\pi\rho(E)|\bra{E} M_{00} \ket{0}|^2$, for $^{12}$C.
The solid lines are the RPA strengths while the dashed lines indicate
the unperturbed ones.
The same smoothing parameter $\Gamma=1$ MeV is used for three
different calculations, (a) BBC (Box boundary condition),
(b) ABC, and (c) OBC.
See text for details.
}
\end{minipage}
}
\label{fig: GMR}
\end{figure}
The calculated HF ground state of $^{12}$C has an oblate shape.
Ratio of the minor and major axises is about two to three.
We carry out three kinds of linear response calculations
with different boundary conditions;
(1) Box boundary condition (BBC) in a 3D coordinate space of radius $R=18$ fm,
(2) Absorbing boundary condition (ABC) in a space of $R=8$ fm plus
    $\Delta r=10$ fm,
and (3) Outgoing boundary condition (OBC) in a space of $R=8$ fm.
The BBC simply means a vanishing boundary condition,
namely $\psi(r)=0$ at $r>R$ fm.
A mesh spacing is taken as $\Delta x=\Delta y=\Delta z=1$ fm.
In order to produce a smooth response curve with BBC,
we adopt a complex energy
$E+i\Gamma/2$ with $\Gamma=1$ MeV.
First, we apply an external field of $M_{00}=r^2$ to see a monopole response.
Results of the calculation are shown in Fig. \ref{fig: GMR}.
The BBC provides a wrong result even though we have used a large box
of $R=18$ fm.
The result of ABC is very similar to that of OBC, except for
small discrepancies in low-energy part ($E<19$ MeV).
Since the highest occupied single-particle energy is $-13.9$ MeV,
energies of outgoing particles are less than 6 MeV in this energy region.
Actually, using $\Delta r=10$ fm,
it is impossible to satisfy the condition,
Eq. (\ref{good_absorber}),
for these low-energy outgoing particles.
We need a larger model space (larger $\Delta r$) to satisfy the condition.
Eq. (\ref{good_absorber}) is satisfied at higher energy,
$E-13.9\mbox{ MeV} \gtrsim 10$ MeV.
The ABC and OBC calculations seem to indicate
two peaks at $E\approx 20$ MeV
(FWHM$\approx 5$ MeV) and at $E\approx 30$ MeV (FWHM$\sim 15$ MeV).

\begin{figure}[t]
\epsfxsize=0.8\textwidth
\centerline{\epsfbox{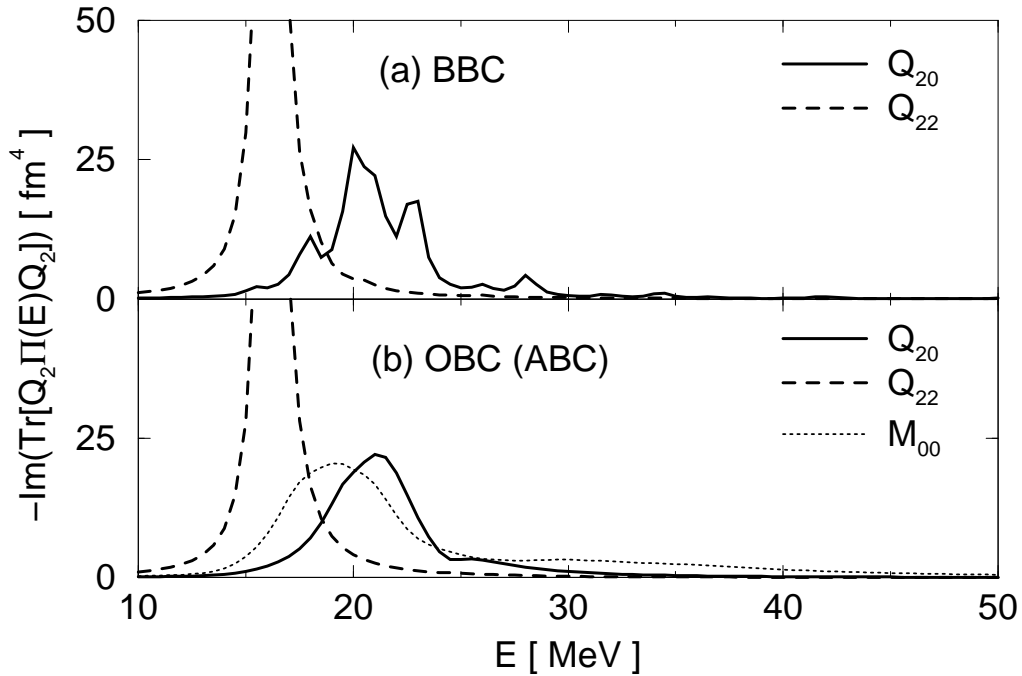}}
\vspace*{-10pt}
\centerline{
\begin{minipage}{0.8\textwidth}
\caption{
Quadrupole strength distribution for $^{12}$C.
The solid lines are the $K=0$ quadrupole strength,
 $\pi\rho(E)|\bra{E}Q_{20}\ket{0}|^2$,
while the dashed lines indicate the $K=2$,
 $\pi\rho(E) |\bra{E}Q_{22}\ket{0}|^2$.
A panel (a) is the result of calculation with BBC,
and (b) is the one with OBC.
The ABC gives the result almost identical to (b).
The smoothing parameter $\Gamma=1$ MeV is used.
The dotted line in a panel (b) indicates a monopole strength distribution
in arbitrary units.
}
\end{minipage}
}
\label{fig: GQR}
\end{figure}
Next, let us discuss isoscalar giant quadrupole resonances (GQR).
The results are shown in Fig. \ref{fig: GQR}.
For GQR, deformation splitting is well-known.
The $K=2$ quadrupole strength is well localized around 16 MeV,
for which the BBC calculation well reproduces the OBC result.
On the other hand, $K=0$ quadrupole strength is located at higher energy
and possesses a wider width.
The single peak at $E\approx 21$ MeV splits into three peaks in the
BBC calculation, although the average energy over the three peaks
becomes a correct value.
Fig. \ref{fig: GQR} (b) also represents the monopole strength
(dotted line).
The results seem to suggest a deformation mixing between
the monopole and $K=0$ quadrupole excitations around $E\sim 20$ MeV.

\section{Conclusions}

Outgoing boundary condition (OBC) is one of the most important ingredients
to treat the continuum.
In order to settle the OBC in the calculation on the 3D mesh space,
we have presented two methods.
The exact treatment can be done by using a Green's function technique,
making resort to an iterative numerical procedure to solve
Eq. (\ref{Psi_eq}).
Another method is to use an absorbing boundary potential.
Carefully choosing the complex potential,
we may simulate the OBC with the same procedure as we do for bound states.
We show some test applications to isoscalar giant resonances
with the BKN Hamiltonian.
The conventional box boundary condition may give wrong results for the
continuum response even if adopting a large box ($R\sim 20$ fm).
The two methods are complementary in a sense that the first method
treats the OBC exactly and the second becomes a very efficient method
for high-energy outgoing particles.
We are preparing applications of present methods
to response and reactions in drip-line nuclei.


\begin{thebibliography}{99}
\bibitem{Flo78}
H.~Flocard, S.~E.~Koonin, and M.~S.~Weiss, \Journal{\PRC}{17}{1682}{1978};
 K.~T.~R.~Davies, H.~Flocard, S.~Krieger, and M.~S.~Weiss,
 \Journal{\NPA}{342}{111}{1980}.
\bibitem{SB75}
S.~Shlomo and G.~Bertsch, \Journal{\NPA}{A243}{507}{1975}.
\bibitem{GS81}
N.~van~Giai and H.~Sagawa, \Journal{\NPA}{A371}{1}{1981}.
\bibitem{HSZ96}
I.~Hamamoto, H.~Sagawa, and X.~Z.~Zhang, \Journal{\PRC}{ 53}{765}{1996};
I.~Hamamoto and H.~Sagawa, \Journal{\PRC}{53}{R1492}{1996};
\Journal{\PRC}{54}{2369}{1996}.
\bibitem{NY01}
T.~Nakatsukasa and K.~Yabana, \Journal{\JCP}{114}{2550}{2001}.
\bibitem{FW71}
A.~L.~Fetter and J.~D.~Walecka, {\em Quantum Theory of Many-Body Systems}
 (McGraw-Hill, New York 1971).
\bibitem{HM78}
I.~Hamamoto and B.~R.~Mottelson, \Journal{\JPSJS}{44}{368}{1978}.
\bibitem{KK86}
R.~Kosloff and D.~Kosloff, \Journal{\JComP}{63}{363}{1986}.
\bibitem{Chi91}
M.~S.~Child, \Journal{\MP}{72}{89}{1991}.
\bibitem{BKN76} P.~Bonche, S.~Koonin, and J.~W.~Negele,
 \Journal{\PRC}{13}{1226}{1976}.
\end{thebibliography}
\end{document}